\documentclass[fleqn,usenatbib]{mnras}

\usepackage{newtxtext,newtxmath}
\usepackage[T1]{fontenc}

\usepackage{graphicx}	
\usepackage{amsmath}	
\usepackage{capt-of}
\usepackage{cases}

\title[M-Earth Habitability Enhanced by Magma Oceans]{The Role of Magma Oceans in Maintaining Surface Water on Rocky Planets Orbiting M-Dwarfs}

\author[Moore, Cowan, \& Boukar\'e]{
Keavin Moore,$^{1,2}$\thanks{E-mail: keavin.moore@mail.mcgill.ca}
Nicolas B. Cowan,$^{1,2,3}$ Charles-\'Edouard Boukar\'e$^{4}$
\\
$^{1}$Department of Earth \& Planetary Sciences, McGill University, 3450 rue University, Montr\'{e}al, QC H3A 0E8, Canada\\
$^{2}$Trottier Space Institute, McGill University, 3550 rue University, Montr\'{e}al, QC H3A 2A7, Canada\\
$^{3}$Department of Physics, McGill University, 3600 rue University, Montr\'{e}al, QC H3A 2T8, Canada\\
$^{4}$Institut de Physique du Globe de Paris, 1 rue Jussieu, Paris CEDEX 05, 75238, France
}

\date{Accepted XXX. Received YYY; in original form ZZZ}

\pubyear{2023}

\begin{document}
\label{firstpage}
\pagerange{\pageref{firstpage}--\pageref{lastpage}}
\maketitle

\begin{abstract}
Earth-like planets orbiting M-dwarf stars, M-Earths, are currently the best targets to search for signatures of life. Life as we know it requires water. The habitability of M-Earths is jeopardized by water loss to space: high flux from young M-dwarf stars can drive the loss of 3--20 Earth oceans from otherwise habitable planets. We develop a 0-D box model for Earth-mass terrestrial exoplanets, orbiting within the habitable zone, which tracks water loss to space and exchange between reservoirs during an early surface magma ocean phase and the longer deep-water cycling phase. A key feature is the duration of the surface magma ocean, assumed concurrent with the runaway greenhouse. This timescale can discriminate between desiccated planets, planets with desiccated mantles but substantial surface water, and planets with significant water sequestered in the mantle. A longer-lived surface magma ocean helps M-Earths retain water: dissolution of water in the magma provides a barrier against significant loss to space during the earliest, most active stage of the host M-dwarf, depending on the water saturation limit of the magma. Although a short-lived basal magma ocean can be beneficial to surface habitability, a long-lived basal magma ocean may sequester significant water in the mantle at the detriment of surface habitability. We find that magma oceans and deep-water cycling can maintain or recover habitable surface conditions on Earth-like planets at the inner edge of the habitable zone around late M-dwarf stars --- these planets would otherwise be desiccated if they form with less than ${\sim}$10 terrestrial oceans of water.
\end{abstract}


\section{Introduction}\label{sec:intro}

\subsection{Habitability}

Planetary habitability requires persistent liquid surface water. The habitable zone around a host star is classically defined as the orbital distance where the incoming stellar flux permits the existence of liquid surface water, assuming an Earth-like atmosphere and silicate weathering thermostat \citep{kasting93, abe93, kopparapu13}, with the runaway greenhouse defining the inner edge and freeze-out of carbon dioxide the outer edge.

The surface water of an Earth-like planet orbiting in the habitable zone of an M-dwarf star is endangered due to flaring and X-ray and extreme ultraviolet radiation emitted by the host star, especially early on when the M-dwarf is most active. Surface temperatures are exceedingly hot and water is readily available in the upper atmosphere to be photodissociated and lost to space (e.g., \citealt{wordsworth13, wordsworth14, luger15, bolmont17, fleming20}).  

M-Earths are common in the Galaxy (e.g., \citealt{dressing15, sabotta21}), motivating theoretical studies of their habitability using predictive coupled models including loss to space. Depending on the space environment and planetary processes, M-Earths could hold onto their surface water throughout their lifetime, recover habitable conditions through degassing of water sequestered within the mantle \citep{moore20}, or become desiccated by the intense irradiation. 

\subsection{The Runaway Greenhouse Phase}

Earth-like planets rapidly form within the habitable zone of M-dwarf stars on Myr timescales (e.g., \citealt{raymond07, lissauer07, ribas14}) or migrate into the habitable zone (e.g, \citealt{petrovich15a, petrovich15b}). Because of the extended, highly-active pre-main sequence phase of its host star, an M-Earth will be highly irradiated following its formation. High surface temperatures on a young M-Earth induce a moist greenhouse, and the high water vapour mixing ratio in the upper atmosphere permits the loss of water to space \citep{kasting93}. If it absorbs too much shortwave radiation then the M-Earth will enter a runaway greenhouse phase. 

Classically, the runaway greenhouse occurs when the incoming irradiation exceeds a critical value; above this value, the moist atmosphere cannot emit sufficient longwave radiation. The oceans completely evaporate into an insulating steam atmosphere, and the greenhouse effect causes surface temperatures to ``run away'' \citep{ingersoll69, kasting88, nakajima92, goldblatt13, goldblatt16}. The length of the runaway greenhouse phase will depend not only on the orbital distance and amount of water in the atmosphere, but also the partial pressure of carbon dioxide and rock vapour (due to the very high temperatures), which cause a blanketing effect. Importantly, since all surface water evaporates into the atmosphere during a runaway greenhouse, the planet is susceptible to lose substantial water to space.

\subsection{Water Loss Regimes}

Water may be lost to space from an M-Earth if it is present above the exobase, or driven by a strong hydrodynamic wind originating deeper in the atmosphere. The loss of water to space from M-Earths is driven by X-ray and extreme ultraviolet (``XUV''; 0.1--120 nm) radiation from the host M-dwarf, which decreases with time \citep{ribas05, luger15}. XUV radiation can photodissociate water molecules and drive their escape to space; for our study, this means the escape of hydrogen atoms as we do not explicitly track atmospheric oxygen. Since the amount of atmospheric water vapour increases with increasing temperature, a hotter planet will lose more water to space. The total amount of water lost may be limited by either incoming energy, or by vertical diffusion of molecules through the atmosphere.

We posit that the loss to space is energy-limited \citep{watson81, erkaev07} when the planet is in a runaway greenhouse. For energy-limited escape, a fixed fraction of the incoming XUV flux, typically 1--10\% (e.g., \citealt{ercolano10, barnes16, lopez17}), powers the photodissociation of water and drives its loss to space. Following the end of the runaway greenhouse phase, once temperatures become more modest, the loss rate is the lower of energy-limited and diffusion-limited. Diffusion-limited escape describes a limit in the upwards diffusion of water towards the exobase \citep{hunten73}; effectively, the diffusion limit corresponds to the vertical diffusion of hydrogen atoms through a background atmosphere of, e.g., oxygen or nitrogen. The diffusion limit applies to minor atmospheric species and thus would not be relevant in a steam atmosphere; however, assuming our model atmosphere consists of water, hydrogen, and oxygen, we can treat the diffusion limit as hydrogen diffusing vertically through an oxygen background. 

\subsection{Surface \& Basal Magma Oceans}

Surface temperatures of ${\sim}$1500--1800 K are expected during the runaway greenhouse phase \citep{kasting88, goldblatt13, goldblatt16}: a steam atmosphere overlies a molten surface --- a surface magma ocean (MO) --- which gradually solidifies from the bottom up over ${\sim}$10s--100s of Myr (e.g., \citealt{elkins08, hamano13, lebrun13}), although this may be extended due to tidal heating, enhanced irradiation, and additional greenhouse gases. The adiabat within the cooling MO crosses the liquidus and solidus. Hence, there are three distinct water reservoirs during the surface magma ocean phase: the steam atmosphere, the shrinking magma ocean, and the growing, solidifying mantle. The partitioning of water during this earliest stage of an M-Earth is critical in determining its water inventory, and its habitability, throughout the planet's lifetime \citep{ikoma18}.

The surface magma ocean is a substantial water reservoir due to the high solubility of water within silicate melts; indeed, for hot planets (such as the young M-Earths investigated in this study), the MO can hold onto enormous amounts of water, up to ${\sim}10$ wt.\% \citep{dorn21}, which is ${\sim}$427 Earth Oceans for an Earth-mass planet. This dissolution of water in the surface magma ocean protects water against the highest irradiation, and against escape to space due to XUV radiation. 

A thick steam atmosphere could extend the surface magma ocean phase, prolonging a reservoir which could sequester substantial oxygen within the mantle \citep{barth21} and preventing the build-up of abiotic O$_2$ in the atmosphere often associated with significant water loss to space (e.g., \citealt{luger15, krissansen21}). Throughout this work, we refer to the early, post-formation magma ocean phase concurrent with the runaway greenhouse as a ``surface magma ocean'', to distinguish it from the later residual ``basal magma ocean'' underlying a solid mantle.

A basal magma ocean (BMO) may form if the mantle solidifies from the middle--out rather than from the bottom--up. A BMO will cool on the order of Gyrs \citep{labrosse07, blanc20}, providing a long-lived water reservoir protected from atmospheric loss. A BMO can also form by density cross-over between solid and liquid silicates \citep{boukare15, boukare17, caracas19}. Regardless of its origin, a BMO will solidify from the top--down, from the base of the solid mantle towards the core-mantle boundary \citep{labrosse07}. A basal magma ocean is likely more common for planets larger than Earth, and has recently been invoked for Venus \citep{orourke20}, Mars \citep{samuel21}, and the Moon \citep{walterova23}.

\subsection{Deep-Water Cycle}

Active plate tectonics provides Earth with a deep-water cycle \citep{mcgovern89}: water is exchanged between surface and mantle reservoirs on geological timescales, driven by the creation and subduction of lithospheric plates. Water is degassed from mantle to surface through mid-ocean ridge volcanism, corresponding to the ``upwelling'' branch of plate tectonics. In the other direction, water is regassed from surface to mantle through subduction of hydrated oceanic crust, corresponding to the ``downwelling'' branch. Although other tectonic modes exist (e.g., stagnant lid, sluggish convection; \citealt{lenardic18, lourenco20}), we restrict the current study to the plate-tectonic mode.

The paper is structured as follows. In Section \ref{sec:methods_stellar}, we outline the equations governing stellar evolution and the loss of water to space within our model. The methods and equations of our magma ocean and deep-water cycling models are then outlined in Section \ref{sec:methods_mo_cycling}, followed by simulation results in Section \ref{sec:results_mo_cycling}. We then discuss our results in Section \ref{sec:discussion}, and our conclusions in Section \ref{sec:conclusions}.

\section{Stellar Evolution \& Water Loss}\label{sec:methods_stellar}

We use stellar models of the bolometric luminosity of M-dwarf stars \citep{baraffe15} to calculate the absorbed flux and surface temperature of a terrestrial planet orbiting in the habitable zone. These stellar models also allow us to calculate the host star's XUV flux as a function of time, and hence estimate the rate of water loss to space in either the energy-limited or diffusion-limited regimes. 

During planet formation, the protoplanetary disk blocks XUV radiation for a few Myr of the ${\sim}$10 Myr formation timescale (e.g., \citealt{barth21}). \citet{ribas14} estimate the lifetimes of protoplanetary disks around low-mass stars to be 4.2--5.8 Myr. Following \citet{ribas14} and \citet{barth21}, we choose to offset the stellar track of \citet{baraffe15} by 5 Myr to allow the circumstellar disk to dissipate and the M-Earth to be fully formed. 

We parameterize the evolution of stellar XUV luminosity over time as \citep{ribas05, luger15}:
\[
    L_{\mathrm{XUV}}(t) = 
\begin{cases}
    f_\mathrm{sat} L_{\mathrm{bol}}(t), & t \leq t_\mathrm{sat}, \\
    f_\mathrm{sat} \left(\frac{t}{t_\mathrm{sat}} \right)^{\beta_\mathrm{XUV}} L_{\mathrm{bol}}(t) ,  & t > t_\mathrm{sat}.
\end{cases}
\]
Here, $t$ corresponds to the time after the gas disk dissipates, $L_\mathrm{bol}(t)$ is the bolometric luminosity interpolated from the stellar tracks of \citet{baraffe15}, $f_\mathrm{sat} = 10^{-3}$ is the saturation fraction, $\beta_{\mathrm{XUV}} = -1.23$, and $t_\mathrm{sat} = 1$ Gyr is the assumed constant saturation time of an M-dwarf. The timescale of spin-down for M-dwarfs --- signalling the end of their highly active phase --- is poorly constrained, and may vary from as low as 600 Myr to ${>}3$ Gyr \citep{pass22}. The bolometric and XUV fluxes at the orbital distance of the M-Earth can then be calculated as $F_\mathrm{bol/XUV} = L_\mathrm{bol/XUV}/(4 \pi a_\mathrm{orb}^2)$, where $a_\mathrm{orb}$ is the M-Earth's orbital distance, a value which we hold fixed throughout each simulation.

\subsection{Parameterization of Atmosphere}

We assume a pure water atmosphere for our model M-Earth. H$_2$O molecules can be broken into two H atoms and one O atom by XUV irradiation. We also adopt a planetary albedo of $A_\mathrm{p} = 0.3$ for all simulated M-Earths, {appropriate for water-dominated atmospheres \citep{kopparapu13}}. Given the calculated absorbed flux, we can then determine the surface temperature, $T_\mathrm{surf}$, assuming a step-wise relation depending on whether or not the planet is in a runaway greenhouse.

If the planet absorbs less than the runaway greenhouse limit for a pure water vapour atmosphere of ${\approx}325$ W/m$^2$ (i.e., the ``water condensation insolation threshold'' of \citealt{turbet21}),  we simply set $T_\mathrm{surf}=293.15$ K $=20^{\circ}$ C. The surface temperature only affects mantle convection, and differences of tens of degrees are negligible. If the planet absorbs more than the runaway greenhouse limit, however, all surface water evaporates into a steam atmosphere, pushing the planet into a runaway greenhouse phase (\citealt{goldblatt13, kopparapu13}). The surface temperature thus runs away to $T_\mathrm{surf}=1800$ K in our model, following \citet{barth21}. We note that this surface temperature arises following the evaporation of 1 Earth Ocean into the atmosphere, although for 8 Earth Oceans, the value is closer to $T_\mathrm{surf}=2500$ K \citep{turbet19}; while we do not explicitly model evaporation, this distinction may be important for future studies.

Loss to space can only occur for particles that are above the exobase, located around 500 km altitude for the Earth \citep{pierrehumbert10} --- unless a hydrodynamic wind is present, which is beyond the scope of this study. Within our model, the loss of water (effectively H atoms) to space will occur in either the energy-limited or diffusion-limited regime.

\subsection{Energy-Limited Escape}

Previously, \citet{luger15} assumed that atmospheric loss to space (either energy-limited or diffusion-limited, tested separately) only occurs during the runaway greenhouse phase. We explicitly account for a steam atmosphere throughout our model. This not only improves its generality, but also allows direct calculation of the amount of water lost to space. We assume water is readily available in the atmosphere due to our relatively long timesteps compared to the short timescales for replenishment by evaporation of surface water.

During runaway greenhouse, we assume that any photodissociated oxygen will dissolve in the surface magma ocean, avoiding the build-up of a significant oxygen background often seen in atmospheric escape studies --- especially \citet{luger15} --- that could hinder the escape of hydrogen to space. Because of the very high surface temperatures, we assume that the loss during runaway greenhouse is always energy-limited.

We adopt the energy-limited escape rate for a pure water atmosphere that can be broken into its constituent hydrogen and oxygen (see \citealt{watson81}, or Equation (2) of \citealt{luger15}):
\begin{equation}\label{eqn:EL_loss}
    \Dot{M}_{\mathrm{EL}} = \frac{\epsilon_{\mathrm{XUV}} \pi F_{\mathrm{XUV}} R_{\mathrm{p}} R_{\mathrm{XUV}}^2}{G M_{\mathrm{p}} K_{\mathrm{tide}}}, 
\end{equation}
where $F_\mathrm{XUV}$ is the XUV flux at the orbital distance of the M-Earth, $R_\mathrm{p}$ the planetary radius, $R_\mathrm{XUV}$ the XUV deposition radius, $M_\mathrm{p}$ the mass of the planet, and $\epsilon_{\mathrm{XUV}}$ is the XUV absorption efficiency. Although \citet{luger15} tested $\epsilon_{\mathrm{XUV}}=$ 0.15--0.3, recent estimates for low-mass planets are closer to 0.1 (e.g., \citealt{owen17}), while \citet{barnes16} tested 0.01--0.15 for Proxima Centauri b; we adopt $\epsilon_{\mathrm{XUV}} = 0.1$ for all host stars. We assume $R_{\mathrm{XUV}} = R_{\mathrm{p}}$ and $K_{\mathrm{tide}} = 1$, since it is of order unity, for simplicity. The former assumption was also made by \citet{luger15}, and may lead to an underestimate of the true energy-limited escape rate \citep{krenn21}. However, \citet{krenn21} also note that $\epsilon_{\mathrm{XUV}} = 0.1$ may in fact overestimate the energy-limited escape rates; thus, our chosen parameters seem justified. Note that adopting $\epsilon_\mathrm{XUV}=0.01$ would reduce the amount of water lost through energy-limited escape by an order-of-magnitude, predictably improving water retention \citep{lopez17}.

\subsection{Diffusion-Limited Escape}

We adopt the diffusion-limited escape parameterization of \citet{luger15}, specifically their Equation (13) (see also \citealt{walker77}, pg. 164). The diffusion-limited escape mass flux of hydrogen atoms is:
\begin{equation}
    \Dot{M}_{\mathrm{DL}} = m_{\mathrm{H}} \pi R_{\mathrm{p}}^2 \frac{b g (m_{\mathrm{O}} - m_{\mathrm{H}})}{k_{\mathrm{B}} T_\mathrm{therm} (1 + X_{\mathrm{O}}/X_{\mathrm{H}})}.
\end{equation}
Here, $m_\mathrm{H}$ and $m_\mathrm{O}$ are the atomic masses of hydrogen and oxygen, respectively, $b = 4.8 \times 10^{17}(T_\mathrm{therm})^{0.75}$ cm$^{-1}$ s$^{-1}$ is the binary diffusion coefficient for hydrogen and oxygen, and $X_\mathrm{O}/X_\mathrm{H} = 1/2$ presumes two hydrogen atoms for every one oxygen atom when water is photodissociated. This assumes that when two hydrogen atoms escape, the corresponding single oxygen atom either escapes as well, or is sequestered through reaction with the solid or molten surface, so that oxygen does not build up in the atmosphere. Finally, we assume a constant thermospheric temperature of $T_\mathrm{therm}=400$ K, as did \citet{luger15}. 

\section{Magma Oceans \& Deep-Water Cycling}\label{sec:methods_mo_cycling}

\citet{moore20} presented a coupled water cycling and loss model with a seafloor-pressure-dependent degassing rate and a mantle-temperature-dependent regassing rate based on \citet{komacek16}. We improve that model by accounting for more dependencies in degassing and regassing. First, however, we will outline the thermal evolution during the surface magma ocean and deep-water cycling phases. 

\subsection{Thermal Evolution Equations}

The surface magma ocean (MO) solidifies from the bottom--up. Due to uncertainties in the timing of magma ocean solidification, especially for planets around M-dwarfs with extended runaway greenhouse phases, we parameterize the surface magma ocean thickness at time $t$ as a function of the MO solidification timescale, $\tau_\mathrm{MO}$:
\begin{equation}\label{eqn:MOdepth}
    d_\mathrm{MO}(t) = \beta_\mathrm{MO} d_{\mathrm{MO},0} \left[\exp \left(\frac{-t}{\tau_\mathrm{MO}} + 1 \right) - 1 \right].
\end{equation}
Here, $\beta_\mathrm{MO} = 1/(\exp(1)-1) \approx 0.582$, and $d_{\mathrm{MO},0}$ is the initial MO depth, from surface to core. Since surface magma ocean and runaway greenhouse phases are concurrent, throughout this work we set $\tau_\mathrm{MO} = \tau_\mathrm{RG}$ (see Table \ref{tab:hzrgmaxwaterloss}) for each host star/orbital distance combination.

The depth of the surface magma ocean decreases as it solidifies from the bottom--up; at time $t$, the bottom of the MO will be located at radius $r(t) = R_\mathrm{p} - d_\mathrm{MO}(t)$. We assume the core heat flux is negligible. We use Equation (1) of \citet{elkins08} to estimate the solidus temperature, based on experimental data of Earth, to determine the MO temperature for a given $r(t)$:
\begin{equation}
    T_\mathrm{MO}(r) = -1.16 \times 10^{-7} r^3 + 0.0014 r^2 - 6.382 r + 1.444 \times 10^4.
\end{equation}

We adopt a timestep during MO of $\mathrm{d}t_\mathrm{MO} = \tau_\mathrm{step,MO} = 2000$ yr due to the comparatively short duration of the surface magma ocean/runaway greenhouse phase compared to the 5 Gyr simulation. 
Once the surface magma ocean solidifies, the mantle begins solid-state convection, initiating plate tectonics and the deep-water cycle on the M-Earth, and we adopt a longer timestep of $\mathrm{d}t = \tau_\mathrm{step} = 20,000$ yr for the remainder of the simulation. 

The mantle temperature during deep-water cycling is calculated following \citet{moore20}, now assuming an initial average mantle temperature of $T_{\mathrm{m},0}=3000$ K following surface magma ocean solidification. We improve the radionuclide heating rate, $Q(t)$, using the equation from \citet{schaefer15}:
\begin{equation}
    Q(t) = \rho_\mathrm{m} \sum C_\mathrm{i} H_\mathrm{i} \exp \left[\lambda_\mathrm{i} \left(4.6 \times 10^9 - t \right) \right].
\end{equation}
Here, $\rho_\mathrm{m}$ is the mantle density. We assume nominal bulk silicate Earth (i.e., ${\sim}21$ ppb U contained in the primitive mantle; \citealt{mcdonough95}). We use the same values as \citet{schaefer15}, taken from \citet{schubert01}, for element concentration by mass, $C_\mathrm{i}$, heat production per unit mass, $H_\mathrm{i}$, and decay constants, $\lambda_\mathrm{i}$, for $^{238}$U, $^{235}$U, $^{40}$K, and $^{232}$Th. We again assume the core heat flux is negligible. 

\subsection{Surface Magma Ocean Water Partitioning}\label{sec:MOeqns}

An M-Earth orbiting within the habitable zone of its host star begins in a surface magma ocean (MO) phase, concurrent with a runaway greenhouse, before surface solidification partitions water between distinct surface and mantle reservoirs. 
Magma ocean dynamics are significantly different than solid mantle dynamics (e.g., \citealt{elkins08, hamano13, ikoma18}). We thus begin our M-Earth simulations with a separate parameterization for MO water partitioning. There are three distinct reservoirs for water during this earliest stage of M-Earth evolution: the steam atmosphere, the surface magma ocean (which shrinks over time), and the solidifying mantle (which grows over time). A box model representation of the surface magma ocean stage is shown in Fig.~\ref{fig:model_flowchart}(a).

Our surface magma ocean approximation assumes bottom--up solidification, typical of many magma ocean studies (e.g., \citealt{elkins08}, \citealt{hamano13}, \citealt{lebrun13}, \citealt{schaefer16}). Water is exsolved into a steam atmosphere once the magma ocean becomes saturated. This canonical bottom--up solidification occurs through fractional crystallization and settling, which leads to a growing solid mantle underlying the shrinking MO. This geologically rapid process usually leads to MO solidification in tens of Myr, but may be extended by an insulating steam atmosphere or high bolometric stellar flux. We posit that the magma ocean lasts as long as the runaway greenhouse: $\tau_\mathrm{MO} = \tau_\mathrm{RG}$. The hot runaway greenhouse temperatures will maintain a molten surface, while the end of runaway greenhouse and the corresponding decrease in surface temperature will permit surface solidification.

We assume that the surface magma ocean is composed of olivine and pyroxenes; these nominally anhydrous silicates have solid-melt water partition coefficients of $D=$ 0.002 and 0.02, respectively, and saturate at 800--1500 ppm (0.08--0.15 wt.\%, depending on mantle composition) of water (see Table 1S of \citealt{elkins08}). In their magma ocean model, \citet{hamano13} assume a water partition coefficient of $D=0.0001$. \citet{katz03} adopt $D=0.01$, due to its similar behaviour to cerium, for their upper mantle melting parameterization. Although the partition coefficient is pressure-dependent and will change with depth as the MO solidifies upwards, we adopt a constant solid-liquid partition coefficient for water of $D=0.001$, intermediate between the values of \citet{katz03} and \citet{hamano13}, since our surface magma ocean is assumed to solidify beginning from the core-mantle boundary at much higher pressures than would be present in the upper mantle. 

We adopt an upper limit on solid mantle water capacity of 12 Earth Oceans \citep{cowan14}, corresponding to ${\sim}3458$ ppm. We vary the surface magma ocean water saturation limit from 0.1 wt.\%, or $C_\mathrm{sat}=0.001$, to 10 wt.\% \citep{elkins08} or $C_\mathrm{sat}=0.1$. Decreasing $C_\mathrm{sat}$ leads to earlier degassing of an atmosphere during the MO phase; beginning with a water inventory larger than the saturation limit will cause an atmosphere to be degassed immediately, resulting in atmospheric loss throughout the surface magma ocean period. 

We assume that, for initial water inventories below $C_\mathrm{sat}$, all water is initially dissolved within the surface magma ocean with no overlying atmosphere. In reality there would always be some partial pressure of water in the atmosphere in equilibrium with the molten surface (e.g., \citealt{elkins08}, \citealt{hamano13}), but the mass of that atmosphere would be dwarfed by the water dissolved in the MO. We bracket this behaviour by testing three values of $C_\mathrm{sat}$, which changes the timing of the onset of atmospheric degassing.

For high enough $C_\mathrm{sat}$, due to the large extent of the MO, the initially dissolved water content is far below the saturation limit. As the MO cools and solidifies, the magma becomes more enriched in water; once $C_\mathrm{sat}$ is exceeded, a steam atmosphere is degassed, from which water may then be lost to space. We assume the entire mantle to be initially molten (see Eqn.~\ref{eqn:MOdepth}). As the MO solidifies, water is partitioned between solid mantle and silicate melt according to our adopted $D=0.001$. Using a value of $C_\mathrm{sat}=0.001$, smaller than the MO concentration of all tested initial water inventories, captures the ongoing degassing of a potentially substantial atmosphere for the entirety of the surface magma ocean, providing a better comparison with simulations that directly calculate the partial pressure of atmospheric water during that stage. 

During the surface magma ocean phase, we use a simple analytical model following \citet{boukare19} to track the amount of water in the three reservoirs --- solid mantle, magma ocean, and steam atmosphere --- as a function of the radius of the bottom of the MO, $r(t)$. Throughout the surface magma ocean phase, we assume a constant density for the silicate melt $\rho = 3000$ kg/m$^3$ for mass balance purposes, although we note that the solid mantle will have a slightly higher density of $\rho_\mathrm{m} = 3300$ kg/m$^3$.

The partitioning model used in this study is based on mass conservation written in an integral form (see \citealt{boukare19}). The concentration of water in the solid phase, $C_\mathrm{s}(r)$, is governed by
\begin{equation}
    C_\mathrm{s}(r) = D C_\mathrm{l}(r),
\end{equation}
where $D=0.001$ is the solid-liquid partition coefficient for water, and $C_\mathrm{l}(r)$ the concentration of water in the liquid phase of the melt. 

While the MO is unsaturated, mass balance allows us to determine $C_\mathrm{l}(r)$ using $R_\mathrm{p}$, the core radius, $R_\mathrm{c}$, and $D$:
\begin{equation}
    C_\mathrm{l}(r) = C_0 \left( \frac{R_\mathrm{p}^3 - R_\mathrm{c}^3}{R_\mathrm{p}^3 - r^3} \right)^{1-D}.
\end{equation}
Here, $C_0$ corresponds to the initial concentration of water within the MO, which we vary to account for different amounts of planetary water inventory from planet formation. Water delivery to terrestrial planets is an ongoing debate (e.g., \citealt{obrien18}). We assume that no more water is accreted by the planet during or after MO solidification. The above equation can also be rearranged to determine the radius at which the MO will be saturated, 
\begin{equation}
    R_\mathrm{sat} = \left( R_\mathrm{p}^3 - \frac{R_\mathrm{p}^3 - R_\mathrm{c}^3}{\left( \frac{C_\mathrm{sat}}{C_0} \right)^{\frac{1}{1-D}}} \right)^{1/3},
\end{equation}
where $C_\mathrm{sat}$ is the water saturation limit of the surface magma ocean.

The mass of water in each reservoir (magma ocean, MO; solid mantle, SM; and atmosphere, atm) can then be determined at each radial extent $r(t)$ for a given time. While the mantle is unsaturated ($r \leq R_\mathrm{sat}$) we have:
\begin{equation}
   M_\mathrm{MO}^\mathrm{uns}(r) = C_\mathrm{l}(r) \frac{4 \pi}{3} \rho \left( R_\mathrm{p}^3 - r^3 \right),
\end{equation}
\begin{equation}
\begin{split}
    M_\mathrm{SM}^\mathrm{uns}(r) & = C_0 \frac{4 \pi}{3} \rho \left( R_\mathrm{p}^3 - R_\mathrm{c}^3 \right)^{1-D} \left[ \left(R_\mathrm{p}^3 - R_\mathrm{c}^3 \right)^D - \left(R_\mathrm{p}^3 - r^3 \right)^D \right] \\
    & = M_\mathrm{init} - M_\mathrm{MO}^\mathrm{uns}(r),
\end{split}
\end{equation}
\begin{equation}
    M_\mathrm{atm}^\mathrm{uns}(r) = 0.
\end{equation}

Once the surface magma ocean saturates (i.e., $r > R_\mathrm{sat}$), water is degassed into the atmosphere. The mass of water in each reservoir following MO saturation is: 
\begin{equation}
    M_\mathrm{MO}^\mathrm{sat}(r) = C_\mathrm{sat} \frac{4 \pi}{3} \rho \left( R_\mathrm{p}^3 - r^3 \right),
\end{equation}
\begin{equation}
    M_\mathrm{SM}^\mathrm{sat}(r) = M_\mathrm{SM}^\mathrm{uns}(R_\mathrm{sat}) + D C_\mathrm{sat} \frac{4 \pi}{3} \rho \left( r^3 - R_\mathrm{sat}^3 \right),
\end{equation}
\begin{equation}
    M_\mathrm{atm}^\mathrm{sat}(r) = M_\mathrm{tot} - M_\mathrm{SM}^\mathrm{sat}(r) -  M_\mathrm{MO}^\mathrm{sat}(r) - \min \left[\Dot{M}_\mathrm{EL}, \frac{M_\mathrm{atm}}{\tau_\mathrm{step,MO}} \right],
\end{equation}
where $M_\mathrm{tot}(t)$ is the total water remaining in the system, accounting for all water that has been lost to space at either the energy-limited rate (due to the runaway greenhouse temperatures) or the entire atmospheric water content in a given timestep. Initially $M_\mathrm{tot} = M_\mathrm{init}$, where,
\begin{equation}\label{eqn:Minit}
    M_\mathrm{init} = C_0 \frac{4 \pi}{3} \rho \left( R_\mathrm{p}^3 - R_\mathrm{c}^3 \right).
\end{equation}

Since we treat the initial amount of water on the M-Earth, $M_\mathrm{init} = W_{\mathrm{MO},i}$, as a free parameter, we can rearrange Eqn.~\ref{eqn:Minit} to determine $C_0$ for a given simulation. In cases where $C_0$ > $C_\mathrm{sat}$, we include an additional first step: the water corresponding to $C_\mathrm{sat}$ is placed in the surface magma ocean, while the excess is immediately degassed into a steam atmosphere. This scenario captures atmospheric degassing and loss throughout the lifetime of the surface magma ocean. 

\subsection{Coupled Cycling \& Loss Equations: Model Without a Basal Magma Ocean (Sans BMO)}
 
Once the planet leaves the runaway greenhouse and the surface solidifies, it begins deep-water cycling; the corresponding box model is shown in Fig.~\ref{fig:model_flowchart}(b). As an improvement to \citet{komacek16} and \citet{moore20}, we incorporate seafloor-pressure, $P$, and mantle-temperature, $T_\mathrm{m}$, dependence into both the degassing and regassing rate. Due to our relatively long timesteps ($\tau_\mathrm{step}=20,000$ yr), we assume that surface water is readily available to vertically diffuse and replenish any water lost to space from the upper atmosphere during a timestep; hence, the atmospheric reservoir is essentially a subset of the surface reservoir in our model. 

The degassing rate, $w_\uparrow$, is given by,
\begin{equation}\label{eqn:degas}
    w_{\uparrow}(P,T_\mathrm{m}) = x \rho_{\mathrm{m}} d_{\mathrm{melt}} f_{\mathrm{degas}}(P) f_{\mathrm{melt}}(T_\mathrm{m}),
\end{equation}
where the pressure dependence is contained in the melt degassing efficiency, $f_{\mathrm{degas}}(P)$:
\begin{equation}\label{eqn:f_degas}
    f_{\mathrm{degas}}(P) = \min \left[ f_{\mathrm{degas,\oplus}} \left(\frac{P}{P_\oplus} \right)^{-\mu}, ~1 \right],
\end{equation}
and the temperature dependence is within the melt fraction, $f_{\mathrm{melt}}(T_\mathrm{m})$:
\begin{equation}\label{eqn:f_melt}
    f_{\mathrm{melt}}(T_\mathrm{m}) = \left(\frac{T_\mathrm{m} - (T_{\mathrm{sol,dry}} - Kx^{\gamma})}{T_{\mathrm{liq,dry}} - T_{\mathrm{sol,dry}}} \right)^{\theta}.
\end{equation} \\
The rest of the above variables are as follows: $x$ is the mantle water mass fraction, $\rho_\mathrm{m}$ is the upper mantle density, $d_\mathrm{melt}$ is the mid-ocean ridge melting depth, $P$ is the seafloor pressure, $P_\oplus = 4 \times 10^7$ Pa is the seafloor pressure for Earth, $T_\mathrm{m}$ is the average mantle temperature, $T_{\mathrm{sol,dry}}$ is the dry solidus temperature of the mantle, $K$ and $\gamma$ are empirically determined constants for solidus depression of a wet mantle \citep{katz03}, $\theta$ is an empirically determined exponent \citep{schaefer15}, and $T_{\mathrm{liq,dry}}$ is the dry liquidus temperature of the mantle.

Note that degassing will occur throughout the deep-water cycling simulation provided there is sufficient water in the mantle and the mantle does not cool below the solidus. We nominally set the exponent $\mu = 1$, following \citet{cowan14} and \citet{komacek16}.

In the other cycling direction, the regassing rate includes $T_\mathrm{m}$-dependent partial melting and $P$-dependent water solubility within the melt, both of which are incorporated into the hydrated layer depth, $d_\mathrm{h}(P, T_\mathrm{m})$. The regassing rate, $w_\downarrow$, is given by
\begin{equation}\label{eqn:regas}
    w_{\downarrow}(P,T_\mathrm{m}) = x_{\mathrm{h}} \rho_{\mathrm{c}} \chi_{\mathrm{r}} d_{\mathrm{h}}(P,T_\mathrm{m}),
\end{equation}
where the hydrated layer depth, $d_{\mathrm{h}}(P,T_\mathrm{m})$, is, 
\begin{equation}\label{eqn:d_h}
\begin{split}
    d_{\mathrm{h}}(P,T_\mathrm{m}) = & \min \biggl[ h^{(1-3\beta)} (T_\mathrm{m} - T_{\mathrm{surf}})^{-(1+\beta)} (T_{\mathrm{serp}} - T_{\mathrm{surf}}) \\
    & \times \left(\frac{\eta(T_\mathrm{m},x) \kappa \mathrm{Ra}_{\mathrm{crit}}}{\alpha \rho_{\mathrm{m}} g} \right)^\beta \left(\frac{P}{P_\oplus} \right)^\sigma, ~d_{\mathrm{b}} \biggr].
    \end{split}
\end{equation}
The regassing-related variables are as follows: $x_\mathrm{h}$ is the water mass fraction in the hydrated crust, $\rho_\mathrm{c}$ is the density of the crust, $\chi_\mathrm{r}$ is the regassing/subduction efficiency, $h$ is the mantle thickness (calculated using the relations of \citealt{valencia06} for a 1 $M_\oplus$ terrestrial planet), $\beta = 0.3$ is an empirically determined constant \citep{mcgovern89}, $T_\mathrm{serp}$ is the serpentinization stability temperature, $\eta(T_\mathrm{m},x)$ is the mantle viscosity, $\kappa$ is the mantle thermal diffusivity, $\mathrm{Ra}_\mathrm{crit}=1100$ is the critical Rayleigh number for mantle convection, $\alpha$ is the mantle characteristic thermal expansivity, $g$ is the surface gravity, and $d_\mathrm{b}$ is the thickness of the basaltic crust.

We also nominally set $\sigma = 1$ \citep{cowan14, komacek16}. For consistency with \citet{moore20}, we include the hydrated layer check from \citet{schaefer15} to ensure the hydrated layer does not contain more water than is present on the surface. Many of the aforementioned deep-water cycling variables can be found in Table C1 of \citet{moore20}.

Combining the equations for degassing, regassing, and loss to space, we can define the evolution of mantle water, $W_\mathrm{m}$, and surface water, $W_\mathrm{s}$, over time during the plate-tectonics-driven deep-water cycling of Fig.~\ref{fig:model_flowchart}(b). The change in mantle and surface water inventories over time are,
\begin{equation}
    \frac{\mathrm{d} W_\mathrm{m}}{\mathrm{d}t} = w_\downarrow - w_\uparrow,
\end{equation}
and
\begin{equation}
    \frac{\mathrm{d} W_\mathrm{s}}{\mathrm{d}t} = w_\uparrow - w_\downarrow - \min \left[ \Dot{M}_\mathrm{DL}, \Dot{M}_\mathrm{EL}, \frac{W_\mathrm{s}}{\tau_\mathrm{step}} \right].
\end{equation}

The amount of water lost to space is either the minimum of the energy-limited or diffusion-limited escape rates during a timestep or the total available surface water during that timestep, as the upper atmosphere water replenishment times are very short compared to $\tau_\mathrm{step}$. For the majority of our simulations, once the surface has solidified, the loss to space is diffusion-limited. Depending on the combination of host star and orbital distance, however, it is possible for the loss to again become energy-limited late in the simulation (see, e.g., Fig.~\ref{fig:TOAflux_lossrates_6TO_mid_log}).

Following \citet{moore20}, we make important checks at each timestep. If the mantle temperature cools below the solidus, melt will no longer be present in the boundary layer, and degassing stops, $w_\uparrow=0$; this is less of an issue for the shorter $t=5$ Gyr simulations within the current study. We also set the minimum water in the mantle to be the same as the desiccation limit for the surface, $W_\mathrm{m,min} \approx 10^{-5}$ Earth Oceans, to avoid mantle viscosity going to infinity.

\subsection{Coupled Cycling \& Loss Equations: Basal Magma Ocean Model}

\begin{figure*}
\centering
\includegraphics[width=1.0\textwidth]{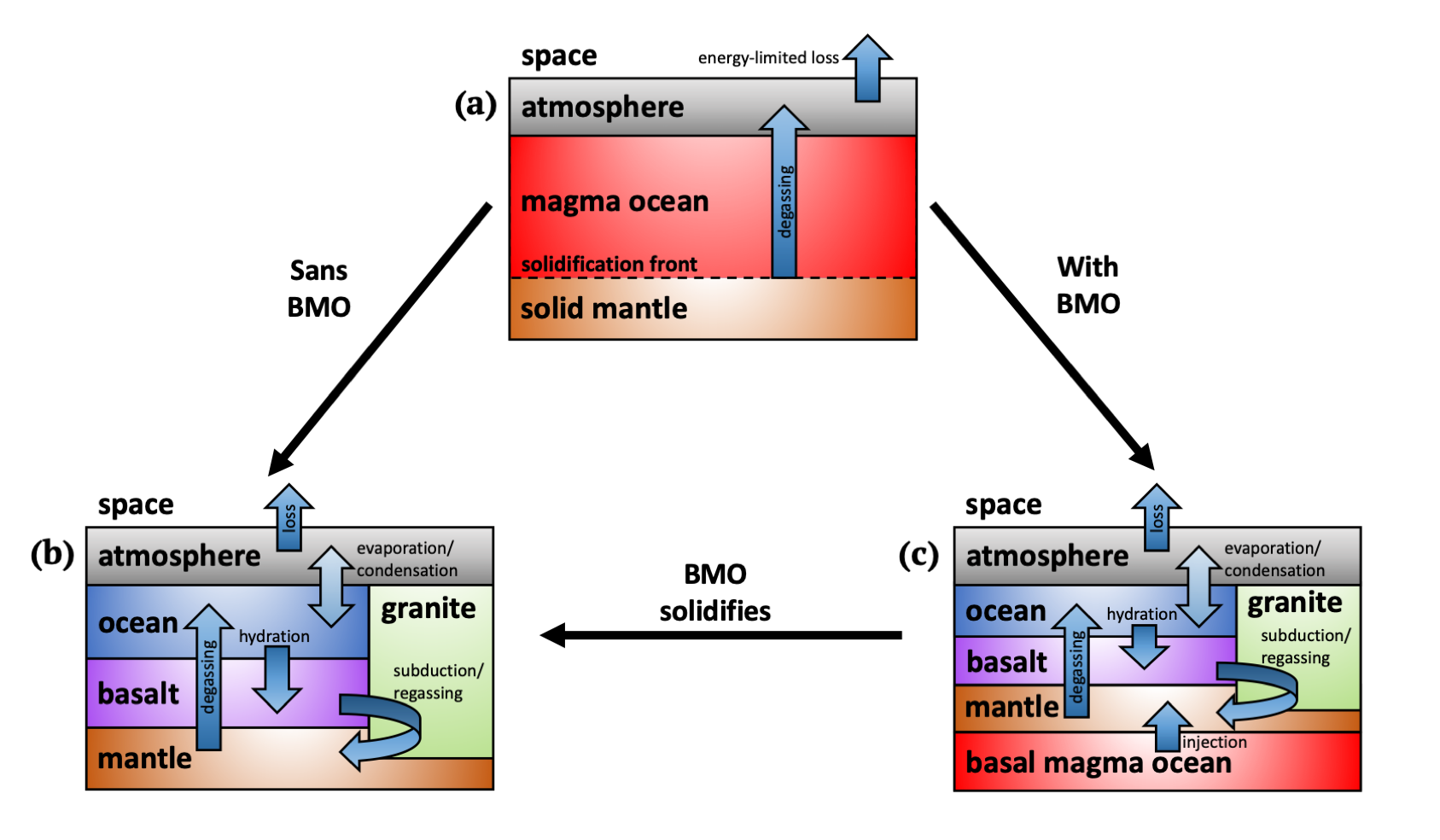}
\caption{\label{fig:model_flowchart} Flowchart illustrating the three possible stages in our box model of M-Earth evolution. (a) Surface magma ocean (MO). We assume bottom--up solidification of the MO lasting as long as the runaway greenhouse ($\tau_\mathrm{MO} = \tau_\mathrm{RG}$). As it solidifies from the bottom--up, the magma ocean eventually becomes saturated with water, and excess water is degassed into a steam atmosphere, from which it may be lost to space through energy-limited escape. (b) Plate-tectonics-driven deep-water cycling including a pure water vapour atmosphere. Water is photodissociated into hydrogen and oxygen high in the atmosphere. Hydrogen may then be lost to space. Water is degassed from mantle to surface through mid-ocean ridge volcanism and regassed from the surface to the mantle through subduction of hydrated oceanic crust. (c) Water cycling in the presence of a basal magma ocean (BMO). After MO solidification, a residual BMO remains below the solid mantle \citep{labrosse07}. While the BMO is present, water may be degassed/regassed, following our deep-water cycling parameterization, or lost to space. Additionally, water is slowly injected into the solid mantle at a constant rate until the BMO completely solidifies. Once the basal magma ocean solidifies at $\tau_\mathrm{BMO}$, the M-Earth evolves from the basal magma ocean model to the deep-water cycling model for the remainder of the simulation. Hence, the two evolutionary pathways are (a)-(b) and (a)-(c)-(b).}
\end{figure*}

Following the surface magma ocean phase and surface solidification, a residual basal magma ocean (BMO) may persist below the solid mantle \citep{labrosse07}. As the BMO solidifies, it will slowly inject water into the overlying mantle. A box model representation of a BMO within an M-Earth during deep-water cycling is shown in Fig.~\ref{fig:model_flowchart}(c). The general idea behind this box model is that a substantial molten reservoir --- within which water is highly soluble --- may exist below the relatively dry mantle following surface magma ocean solidification, and the slow incorporation of this water into the mantle should prevent mantle desiccation.

The two key parameters governing the magma ocean in our model are its extent, which determines water storage, and the solidification timescale. To roughly mimic the behaviour of a BMO, we run our MO simulations using an exceedingly high $D=0.2$ to account for water stored in both the solid mantle and in the BMO; this higher $D$ will result in a significant portion of the water locked into the planetary interior at the end of the MO phase. Once the MO solidifies, the mantle contains the same small amount as the corresponding $D=0.001$ MO simulation, while the remainder --- and the lion's share --- is attributed to the BMO.

Following surface magma ocean solidification, the cycling equation for the surface remains unchanged, while the solid mantle equation now accounts for a slow but constant injection of water until the basal magma ocean solidifies at $t = \tau_\mathrm{BMO}$:
\begin{equation}
    \frac{\mathrm{d} W_\mathrm{m}}{\mathrm{d}t} = w_\downarrow - w_\uparrow + \left(\frac{W_\mathrm{BMO,i}}{\tau_\mathrm{BMO} - \tau_\mathrm{MO}}\right),
\end{equation}
where $W_\mathrm{BMO,i}$ is the initial amount of water in the basal magma ocean at the time of surface solidification. Since a basal magma ocean cools on the order of Gyrs \citep{labrosse07, blanc20}, we adopt $\tau_\mathrm{BMO} = $ 1, 2, or 3 Gyr.

There are two potential evolutionary pathways for each simulated M-Earth, illustrated in Fig.~\ref{fig:model_flowchart}: either our sans basal magma ocean simulations (left), or simulations incorporating a basal magma ocean (right), which solidifies to the deep-water cycling box model.

\subsection{Model Inputs}

The M stellar type spans a wide range of masses and radii. To bracket the potential impact of different host stars on the water evolution of an orbiting M-Earth, we test three host M-dwarf stars using the stellar tracks of \citet{baraffe15}: a Proxima Centauri-like M5 ($0.13 M_\odot$), a smaller M8 ($0.09 M_\odot$), and a larger M1 ($0.50 M_\odot$). We test three fixed orbital distances within the habitable zone (HZ) of each host star, calculated at $t=4.5$ Gyr using the HZ calculator of \citet{kopparapu13}. We can then calculate the duration of runaway greenhouse at this fixed orbital distance for each combination of host star and location within HZ, to be adopted as the duration of the surface magma ocean, i.e., $\tau_\mathrm{MO} = \tau_\mathrm{RG}$; these values can be found in Table \ref{tab:hzrgmaxwaterloss}. For simplicity in our comparisons, we neglect the difference in runaway greenhouse length when varying the atmospheric water content, although we note that complete desiccation of the planet would end the runaway greenhouse.

Recall that we offset the stellar tracks of \citet{baraffe15} by 5 Myr to account for planet formation. For simplicity, throughout this work we assume a constant planetary albedo for all simulated M-Earths of $A_\mathrm{p} = 0.3$ motivated by Earth's present-day albedo, a parameter also assumed constant in the magma ocean simulations of \citet{hamano13}. This albedo is intermediate between that of a water-covered surface, $A_\mathrm{p} =$ 0.05--0.1 \citep{peixoto92} and the albedo of the ``steam atmosphere'' of Venus, $A_\mathrm{p}=0.75$ \citep{schaefer16, barth21}; hence, assuming $A_\mathrm{p} = 0.3$ may overestimate the absorbed radiation during runaway greenhouse.

Predictions for water inventories of terrestrial exoplanets vary wildly (e.g., \citealt{raymond04, raymond07, lissauer07, obrien18, lichtenberg19, kimura22}). Earth's oceans contain ${\sim}1.4 \times 10^{21}$ kg $\equiv 1$ Earth Ocean of water, and the mantle's capacity may be up to 12 Earth Oceans \citep{bercovici03, hirschmann06, cowan14, guimond22}. The water capacity of a planetary mantle will depend on its mineralogy/composition, oxygen fugacity, and planetary mass. \citet{shah21} determined constraints on the internal storage of water in terrestrial planets, estimated to be 0--6 wt.\% (0 to ${\sim}$250 Earth Oceans). Recently, \citet{barth21} found that for initial water inventories ranging from 1--100 Earth Oceans, only 3--5\% (0.03--5 Earth Oceans) will be sequestered within the mantle of TRAPPIST-1e, 1f, and 1g after surface magma ocean solidification.

We test a range of initial water inventories, surface magma ocean water saturation limits, and basal magma ocean lifetimes. We explore the parameter space in Table \ref{tab:param_space} for the first 5 Gyr of an M-Earth's lifetime. We test water inventories from $M_\mathrm{init}=$ 2 to 400 Earth Oceans (initially completely dissolved in the surface magma ocean if possible; otherwise, the excess above saturation is immediately degassed), three surface magma ocean water saturation limits --- $C_\mathrm{sat}=$ 0.001, 0.01, and 0.1 --- and three basal magma ocean lifetimes of $\tau_\mathrm{BMO}=$ 1, 2, and 3 Gyr. 

Recall that for each simulation, we set $\tau_\mathrm{MO} = \tau_\mathrm{RG}$; we avoid explicitly modelling the feedback between water degassing into a (potentially) runaway greenhouse atmosphere and surface magma ocean solidification. Variation in $\tau_\mathrm{RG}$, and hence $\tau_\mathrm{MO}$, is achieved through the combination of host star and orbital distance (see Table \ref{tab:hzrgmaxwaterloss}). Our surface magma ocean solidification timescales roughly correspond to the range found by \citet{schaefer16} (their Fig.~5, for the highly-irradiated GJ 1132b) and \citet{barth21} (their Fig.~2), while \citet{hamano13} note that planets with comparable water budgets to the modern Earth require a few Myr to 100 Myr for surface magma ocean solidification. 

\begin{table*}
    \centering
    \begin{tabular}{c|c|c}
    \hline
        Name & Parameter & Values Tested  \\
    \hline
         Total initial water mass & $M_\mathrm{init}$ [Earth Oceans] & {\bf 2, 4, 6, 8}, 12, 16, 20, 24, 50, 100, 200, 400 \\
         Surface magma ocean (MO) saturation limit & $C_\mathrm{sat}$ & 0.001, {\bf 0.01}, 0.1 \\
         Basal magma ocean (BMO) solidification timescale & $\tau_\mathrm{BMO}$ [Gyr] & 1, {\bf 2}, 3 \\
         Orbital distance/location within HZ & $a_\mathrm{orb}$ & {\bf Inner HZ, Mid HZ, Outer HZ} \\
         Host stellar type & --- & {\bf M8, M5, M1} \\
         Water loss prescription & --- & {\bf $\Dot{M}_\mathrm{EL}$ during MO, $\min \left[\Dot{M}_\mathrm{DL}, \Dot{M}_\mathrm{EL} \right]$ otherwise}; \\
        & & $\Dot{M}_\mathrm{EL}$ throughout \\
    \hline
    \end{tabular}
    \caption{Parameters explored in this study. Water mass is expressed in units of Earth Oceans, $\approx 1.4 \times 10^{21}$ kg, while the value of $a_\mathrm{orb}$ varies depending on the combination of host star and location within the habitable zone. Bold values indicate our nominal parameter space.}
    \label{tab:param_space}
\end{table*}

\section{Results: Coupled Magma Oceans \& Deep-Water Cycling}\label{sec:results_mo_cycling}

\subsection{Temporal Fluxes \& Expected Water Loss}\label{sec:stellar_results}

Our chosen stellar hosts, locations within the habitable zone, and orbital distances are listed in Table \ref{tab:hzrgmaxwaterloss}. Further, we report the duration of the runaway greenhouse (RG) based on absorbed flux, which will later be used as the surface magma ocean duration ($\tau_\mathrm{MO} = \tau_\mathrm{RG}$). We calculate the maximum water that could be lost using the most likely scenario of energy-limited loss during RG and diffusion-limited loss otherwise, along with the maximum if energy-limited loss occurs for the entirety of each simulation. An M-Earth orbiting at the inner edge of the habitable zone loses much more water through energy-limited loss during runaway greenhouse than the rest of the simulation; conversely, at the outer edge of the habitable zone, more water is lost during the long diffusion-limited stage than during the brief runaway greenhouse phase.

\begin{table*}
    \centering
    \begin{tabular}{c|c|c|c|c|c|c|c|c}
    \hline
        Stellar & Stellar & Location Within & Orbital & Runaway Greenhouse & \multicolumn{4}{c}{Maximum Water Loss [Earth Oceans]} \\
        \cline{6-9}
        Type & Mass & Habitable Zone (HZ) & Distance & (RG) Duration, $\tau_\mathrm{RG}$ & EL during RG & DL rest of sim & Total & Max, EL only \\
    \hline
         M8 & $0.09M_\odot$ & Hot, Inner Edge & 0.025 AU & 335 Myr (0.335 Gyr) & 7.72 & 2.82 & 10.5 & 19.7 \\
          & & Middle (Mid) & 0.037 AU & 110 Myr (0.110 Gyr) & 2.18 & 2.95 & 5.13 & 8.64 \\
          & & Cold, Outer Edge & 0.050 AU & 48.2 Myr (0.048 Gyr) & 0.833 & 2.99 & 3.83 & 4.83 \\
    \hline
         M5 & $0.13M_\odot$ & Inner & 0.045 AU & 155 Myr (0.155 Gyr) & 2.92 & 2.92 & 5.85 & 15.8 \\
         & & Mid & 0.067 AU & 44.5 Myr (0.044 Gyr) & 0.713 & 2.99 & 3.71 & 7.09 \\
         & & Outer & 0.089 AU & 17.5 Myr (0.018 Gyr) & 0.235 & 3.01 & 3.25 & 4.01 \\
    \hline
         M1 & $0.50M_\odot$ & Inner & 0.196 AU & 33.8 Myr (0.034 Gyr) & 0.475 & 3.00 & 3.50 & 13.8 \\
         & & Mid & 0.286 AU & 6.59 Myr (0.0066 Gyr) & 0.074 & 3.02 & 3.09 & 6.47 \\
         & & Outer & 0.377 AU & 0.36 Myr (0.00036 Gyr) & 0.003 & 3.02 & 3.02 & 3.73 \\
    \hline
    \end{tabular}
    \caption{Host star spectral classification and mass used in our simulations, from stellar evolution tracks of \citet{baraffe15}. The corresponding location within the habitable zone (HZ) and orbital distance around each star is calculated at $t = 4.5$ Gyr using the HZ calculator of \citet{kopparapu13}. We compute the runaway greenhouse (RG) duration from the time-dependent irradiation, assuming a constant planetary albedo of $A_\mathrm{p}=0.3$. The potential water loss in different regimes is included in the next columns: energy-limited (EL) loss assuming an efficiency of $\epsilon_\mathrm{XUV}=0.1$ during RG, diffusion-limited (DL) loss following the end of RG, and the combined total potential water loss. The final column shows the maximum potential water loss if energy-limited for the entire simulation; due to our parameterization, the maximum water loss when diffusion-limited for the entire simulation is always 3.02 Earth Oceans. Potential water loss is expressed in units of Earth Oceans, $\approx 1.4 \times 10^{21}$ kg. Depending on the details of atmospheric loss, M-Earths can lose 3--20 Earth Ocean of water after 5 Gyr. Note that reducing $\epsilon_\mathrm{XUV}$ by an order-of-magnitude, from 0.1 to 0.01 \citep{lopez17}, would also reduce the amount of water lost through energy-limited escape by an order-of-magnitude.}
    \label{tab:hzrgmaxwaterloss}
\end{table*}

\begin{figure}
\centering
\includegraphics[width=0.48\textwidth]{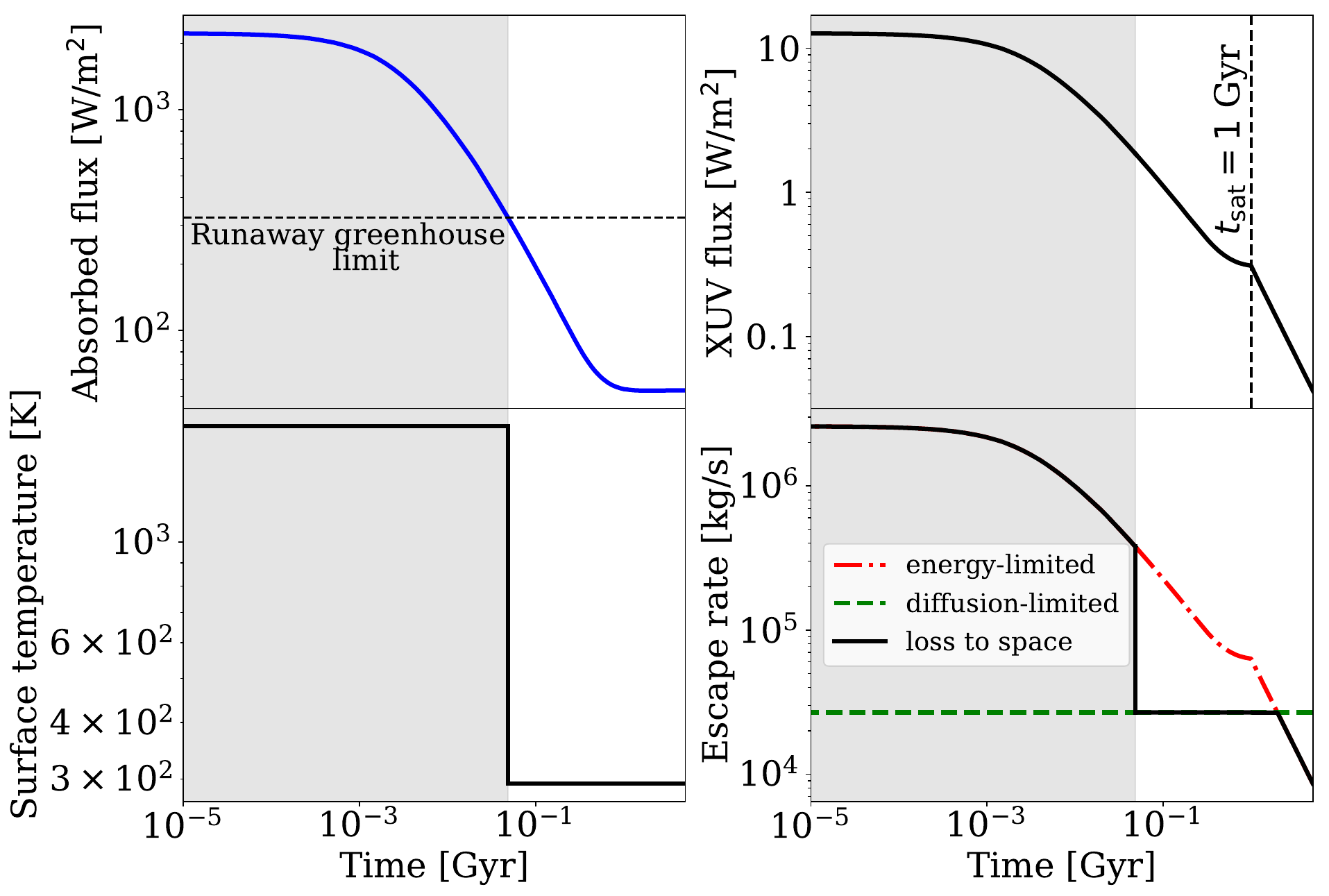}
\caption{\label{fig:TOAflux_lossrates_6TO_mid_log} Evolution of the absorbed flux assuming a constant $A_\mathrm{p}=0.3$ (top left), surface temperature (bottom left), X-ray and extreme ultraviolet flux (top right), and atmospheric escape rate (bottom right). This figure corresponds to an Earth-like planet orbiting at the outer edge of the habitable zone of an M8 host star. During the runaway greenhouse phase (shaded grey), loss to space is energy-limited with an efficiency of $\epsilon_\mathrm{XUV}=0.1$, and $T_\mathrm{surf}$ is held constant at 1800 K. Once the M-Earth exits the runaway greenhouse, loss to space becomes diffusion-limited, and $T_\mathrm{surf}$ is fixed at 293.15 K. The loss again becomes energy-limited (i.e., the lower of diffusion- and energy-limited) near the end of the simulation. The cusp in $F_\mathrm{XUV}$ and the energy-limited escape rate corresponds to our adopted stellar XUV saturation timescale of $t_\mathrm{sat}=1$ Gyr. Although loss rates are much greater during runaway greenhouse, the integrated loss for many planets is dominated by diffusion-limited loss during post-runaway greenhouse, especially for higher magma saturation limits $C_\mathrm{sat}$ which leads to later degassing of an atmosphere.} 
\end{figure}

Fig.~\ref{fig:TOAflux_lossrates_6TO_mid_log} shows the evolution of radiative fluxes, surface temperature, and atmospheric loss rate for an Earth-like planet orbiting at the outer edge of the habitable zone (Outer HZ) of an M8 host star. The runaway greenhouse phase is shaded grey. Once the absorbed flux falls below the runaway greenhouse limit of 325 W/m$^2$ \citep{turbet21}, the surface temperature decreases from the runaway greenhouse limit of $T_\mathrm{surf}=1800$ K to a temperate $T_\mathrm{surf}=293.15$ K. The evolution of the rate of loss to space is clear in the lower right panel: during runaway greenhouse, the loss to space is energy-limited, before switching to diffusion-limited when the runaway greenhouse ends, and once again becoming energy-limited (the lower of the two escape rates) near the end of the simulation.

\subsection{Water Evolution on Specific Planets}

\begin{figure}
\centering
\includegraphics[width=0.45\textwidth]{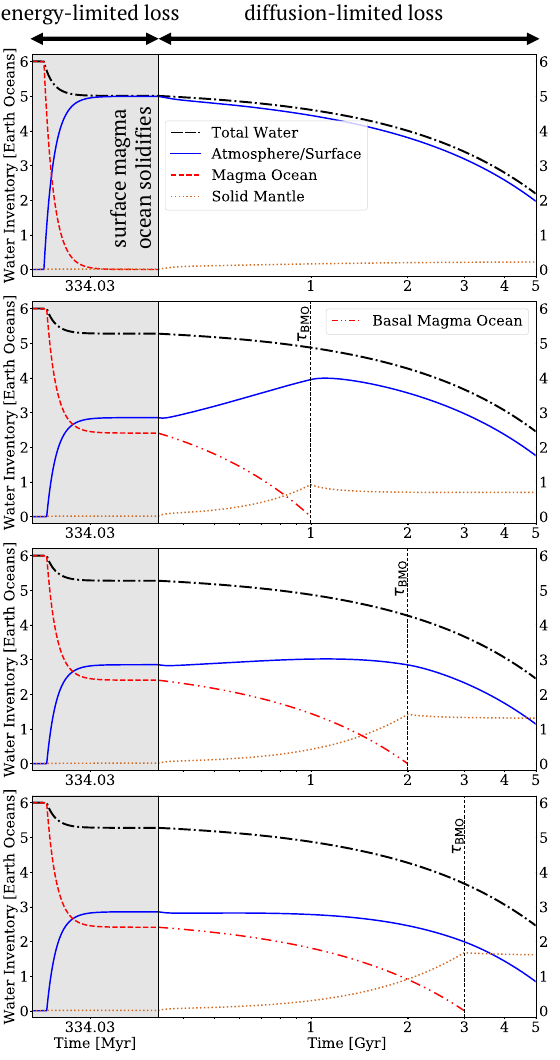}
\caption{\label{fig:MO_cycling_6TO_RG_M8_diffBMOs} Water evolution for an M-Earth orbiting at the inner edge of the habitable zone (Inner HZ) of an M8 host star. The planet is initialized with 6 Earth Oceans; the surface magma ocean water saturation limit is $C_\mathrm{sat}=0.01$, and the surface magma ocean/runaway greenhouse phase 
(left part of the figure, shaded grey and plotted on a reverse log scale) lasts $\tau_\mathrm{MO} = \tau_\mathrm{RG} \approx $ 335 Myr. The topmost panel corresponds to the model without a basal magma ocean, while the following three display results for different basal magma ocean (BMO) lifetimes, $\tau_\mathrm{BMO}$. In the simulation sans BMO (top), most water is in the atmosphere/surface reservoir following surface magma ocean solidification, where it is susceptible to be lost to space. In contrast, substantial water remains sequestered within a BMO while only ${\sim}60$\% of the total water is degassed into the atmosphere. Longer-lived basal magma oceans lead to slower injection and more water trapped in the cooling mantle; indeed, for $\tau_\mathrm{BMO} = $ 1 Gyr, the injection into the relatively hot mantle allows the surface water inventory to briefly grow, while $\tau_\mathrm{BMO} = $ 3 Gyr results in more water within the mantle than at the surface. Following BMO solidification, water continues to be lost from the surface at the diffusion-limited rate; by 5 Gyr, the planet retains water and remains habitable in all scenarios. The presence of a basal magma ocean improves water retention, but at the detriment of surface habitability: the water sequestered in the basal magma ocean tends to remain in the mantle.}
\end{figure}

We present in Fig.~\ref{fig:MO_cycling_6TO_RG_M8_diffBMOs} a representative example of the temporal evolution of water inventories for the first 5 Gyrs of an Earth-like planet orbiting an M8 host star at the Inner HZ with an initial water inventory of 6 Earth Oceans; this figure includes simulations without and with a basal magma ocean for various basal magma ocean lifetimes. The left part of the figure (grey area, and plotted on a reverse-log scale) corresponds to the concurrent surface magma ocean/runaway greenhouse phase, while the right part (white area) corresponds to the longer deep-water cycling phase.

In the sans basal magma ocean simulation (top of Fig.~\ref{fig:MO_cycling_6TO_RG_M8_diffBMOs}), an atmosphere is degassed during the surface magma ocean (MO) phase once the shrinking MO reaches saturation. Water is then lost from the atmosphere at the energy-limited rate. By $\tau_\mathrm{MO}$, the time of surface solidification, ${\sim}1$ Earth Ocean has been lost to space; most of the remaining ${\sim}$5 Earth Oceans ends up in the surface/atmosphere reservoir. A small amount of water remains in the solid mantle throughout the simulation, governed by our water partition coefficient $D=$ 0.001. Water is lost at the diffusion-limited rate following surface solidification, and the M-Earth ends the 5 Gyr simulation in a habitable regime with ${\sim}2$ Earth Oceans of surface water. Note that following surface solidification, water loss to space remains diffusion-limited for the rest of the simulation, as opposed to the brief switch back to energy-limited shown in Fig.~\ref{fig:TOAflux_lossrates_6TO_mid_log}.

A basal magma ocean (BMO) does not significantly alter the habitability prospects of this M-Earth (bottom three panels of Fig.~\ref{fig:MO_cycling_6TO_RG_M8_diffBMOs}) regardless of BMO lifetime, $\tau_\mathrm{BMO}$. Following MO solidification and the energy-limited loss of ${\sim}0.8$ Earth Oceans to space, only ${\sim}2.8$ Earth Oceans of water has been degassed into the atmosphere, while ${\sim}2.4$ Earth Oceans remain locked within the BMO. Water is injected into the overlying solid mantle by the slowly solidifying BMO at a constant rate governed by $\tau_\mathrm{BMO}$. Although the total final water inventory is the same for all basal magma ocean simulations in Fig.~\ref{fig:MO_cycling_6TO_RG_M8_diffBMOs}, the temporal evolution and final partitioning of water between reservoirs differs. 

A shorter $\tau_\mathrm{BMO}$ leads to injection of water into a still relatively hot mantle, which permits the surface inventory to increase, even in the face of energy-limited loss, until the basal magma ocean disappears. Increasing $\tau_\mathrm{BMO}$ leads to slower water injection into a cooler mantle which does not release the water to the surface; as a result, mantle water inventory increases with increasing BMO lifetime. Indeed, the shortest $\tau_\mathrm{BMO}$ leads to substantial surface water and some mantle water, while the longest $\tau_\mathrm{BMO}$ results in ${\sim}$50\% more water sequestered within the mantle than is present on the surface. 

Regardless of whether a basal magma ocean is included in the simulation, by 5 Gyr this particular M-Earth is in a habitable regime with ${\sim}$1--2 Earth Oceans on its surface. This is all the more impressive because the planet was expected to lose 10.5 Earth Oceans (Table \ref{tab:hzrgmaxwaterloss}).  Fig.~\ref{fig:MO_cycling_6TO_RG_M8_diffBMOs} therefore demonstrates the ability of a long-lived magma ocean, along with a deep-water cycle, to maintain the habitability of an M-Earth in the face of predicted significant loss to space based solely on stellar evolution.

Although these simulations present a clear benefit of a coupled magma ocean and deep-water cycle, the basal magma ocean itself does not seem to significantly aid in maintaining habitable conditions, with the final total water inventory only slightly higher than in the simulations without a basal magma ocean. Many BMO simulations in our tested parameter space end up with a significant portion of water trapped within the solid mantle and little to none on the surface when compared to the same scenario in the simulations sans BMO. This reiterates the previous results of, e.g., \citet{schaefer15} and \citet{korenaga17}: an aging planet soaks up water in the mantle, sometimes to the detriment of surface habitability.

Depending on $\tau_\mathrm{BMO}$, the basal magma ocean injection rate could be similar to the diffusion-limited loss rate. If atmospheric loss is predominantly diffusion-limited for the majority of the simulation, as our model inherently assumes, then a long-lived basal magma ocean helps keep the mantle hydrated but does not keep the surface habitable.

\begin{figure}
\centering
\includegraphics[width=0.45\textwidth]{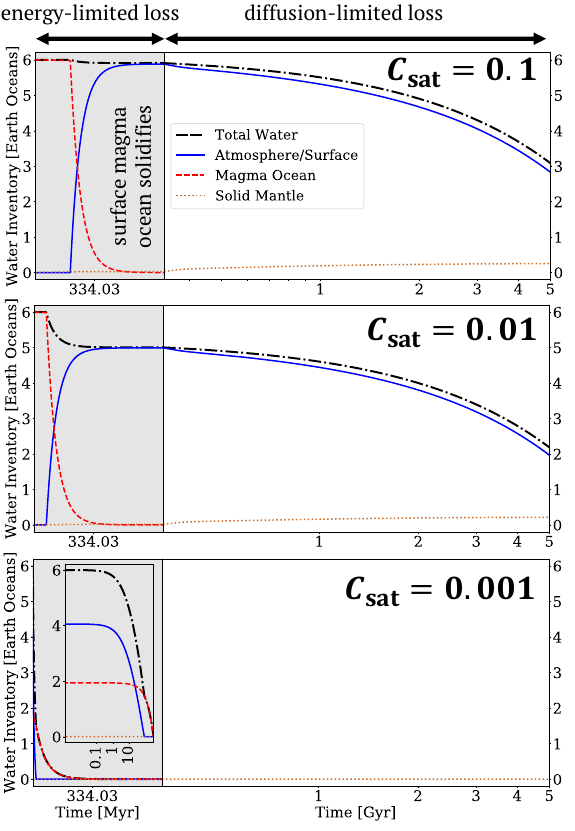}
\caption{\label{fig:MO_cycling_6TO_RG_M8_diffCsat} Same as the case in the top panel of Fig.~\ref{fig:MO_cycling_6TO_RG_M8_diffBMOs}, but now for different surface magma ocean (MO) water saturation limits, $C_\mathrm{sat}$. Decreasing $C_\mathrm{sat}$ leads to earlier degassing of an atmosphere during MO, meaning energy-limited loss of water to space begins earlier as well. This is clear when comparing the top two panels: ${\sim}$1 Earth Ocean more water is lost when $C_\mathrm{sat}$ is decreased from 0.1 to 0.01. For $C_\mathrm{sat} = 0.001$ (bottom panel), an atmosphere is degassed immediately since the saturation limit is below the initial water inventory. Because the evolution of water inventories is difficult to see on the reverse-log scale, we include an inset panel during the MO phase, which makes it clear that immediate degassing and ongoing energy-limited loss with $\epsilon_\mathrm{XUV}=0.1$ rapidly leads to desiccation of the atmosphere/surface, and a very small amount of water locked within the solid mantle. For sufficiently high water saturation limits, an M-Earth is able to survive desiccation, despite the expected 10.5 Earth Oceans of loss (see Table \ref{tab:hzrgmaxwaterloss}), through dissolution of water within the surface magma ocean, which can protect it from loss to space for long periods.}
\end{figure}

Before moving on, we include Fig.~\ref{fig:MO_cycling_6TO_RG_M8_diffCsat} to display the evolution of water inventories when varying the surface magma ocean water saturation limit $C_\mathrm{sat}$. The top two panels show similar behaviour for $C_\mathrm{sat}$ = 0.1 and 0.01, albeit with different amounts of water loss to space. This is expected, since decreasing $C_\mathrm{sat}$ leads to earlier atmospheric degassing and hence earlier energy-limited loss, which then occurs for the remainder of the MO phase. 

The bottom panel of Fig.~\ref{fig:MO_cycling_6TO_RG_M8_diffCsat} shows $C_\mathrm{sat}$ = 0.001, a value which is below the initial water inventory of 6 Earth Oceans. Because $C_0 < C_\mathrm{sat}$, an atmosphere is immediately degassed, and energy-limited loss is ongoing for the entire MO phase. Since atmospheric desiccation is very rapid, the inset panel makes the evolution of the various water inventories much clearer: the planet nearly becomes completely desiccated, save for a small amount of water sequestered in the solid mantle. Hence, Fig.~\ref{fig:MO_cycling_6TO_RG_M8_diffCsat} shows that, depending on $C_\mathrm{sat}$, an M-Earth which may be expected to lose substantial water to space can be saved through dissolution of water within the surface magma ocean for extended periods. We further explore this behaviour as we explore parameter space in the following section.

\subsection{Parameter Space Exploration}

Next, we include results for a region of parameter space explored for a given host star. Specifically, we can represent the evolution of water inventory for each combination of initial water inventory $M_{\mathrm{init}}$ and location within the habitable zone. 

The parameter space results for M1, M5, and M8 host stars (not shown) look nearly identical between their corresponding simulations without and with a basal magma ocean (BMO), albeit with final water inventories differing by a few percent. This supports the result of Fig.~\ref{fig:MO_cycling_6TO_RG_M8_diffBMOs}: regardless of host star, a BMO does not significantly alter the habitability prospects when surface water is used as a proxy for habitability. Indeed, a longer BMO can be detrimental to surface habitability as water becomes trapped in the mantle below a desiccated surface. For all host stars, M-Earths initiated with 2 Earth Oceans become completely desiccated, while M-Earths with more initial water are often able to avoid or recover from desiccation, depending on their orbital distance and surface magma ocean/runaway greenhouse duration. Interestingly, for the M5 host star, $C_\mathrm{sat}=0.01$, and an initial water content of 4 Earth Oceans at the Inner HZ leads to a desiccated surface for the BMO simulation, but the sans BMO simulation survives in a habitable regime; this also occurs for 6 Earth Oceans when $C_\mathrm{sat}=0.001$, and for the M1 host star for 4 Earth Oceans and $C_\mathrm{sat}=0.1$. This is due to water becoming trapped in the solid mantle as water is injected into the mantle from the basal magma ocean at a constant rate. 

Further, the results are almost the same for surface magma ocean (MO) water saturation limits of $C_\mathrm{sat}=$ 0.1 and 0.01. For these highest tested values of $C_\mathrm{sat}$, water is protected from extensive loss to space during the MO phase, as atmospheric degassing is delayed until very close to the end of the MO phase due to the high solubility of water within the magma. For $C_\mathrm{sat}=0.01$, substantial water becomes locked in the mantle during the deep-water cycling phase, which can result in a desiccated surface above a hydrated mantle.

For $C_\mathrm{sat}=0.001$, however, planets lose significantly more water to space, with more planets becoming desiccated due to the earlier atmospheric degassing and ongoing energy-limited loss during the surface magma ocean/runaway greenhouse phase. This low magma ocean saturation limit is below the concentration of all tested $M_\mathrm{init}$, meaning the surface magma ocean begins saturated with water and thus an atmosphere is immediately degassed. The M1 host star simulations are the most similar for different values of $C_\mathrm{sat}$, owing to the comparatively shorter runaway greenhouse phases experienced by the orbiting planets. 

\begin{figure}
\centering
\includegraphics[width=0.28\textwidth]{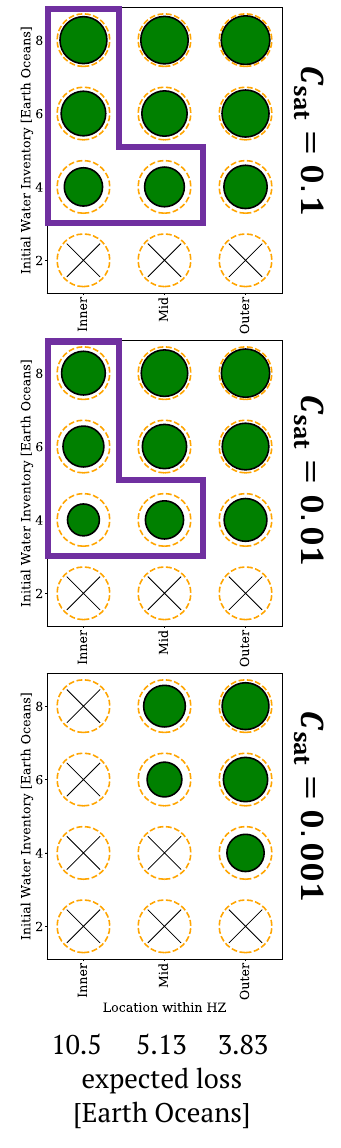}
\caption{\label{fig:grid_m8}Parameter space results an Earth-like planet without a basal magma ocean orbiting an M8 host star. Each panel corresponds to a different surface magma ocean (MO) water saturation limit, $C_\mathrm{sat}$. Within each panel, the initial water inventory is illustrated by a dashed orange circle, while the final water inventory is represented by a green-filled circle, scaled relative to the dashed orange circle to represent the fraction of water lost (a black X denotes a desiccated surface). The expected loss shown along the bottom of the figure is taken from Table \ref{tab:hzrgmaxwaterloss}. Decreasing $C_\mathrm{sat}$ leads to earlier degassing of an atmosphere during the surface magma ocean phase, and hence more extensive energy-limited loss. Indeed, the bottom panel corresponds to a water saturation limit \textit{below} all tested initial water inventories; since the MO begins saturated, a substantial atmosphere is immediately degassed, and atmospheric degassing is ongoing throughout the MO phase. Purple boxes indicate scenarios where the M-Earth started with less water than it was expected to lose (Table \ref{tab:hzrgmaxwaterloss}), but nonetheless ended with significant surface water: these survivors are a testament to the ability of a long-lived magma ocean to protect a planet's water from loss to space.}
\end{figure}

Fig.~\ref{fig:grid_m8} presents the parameter space results, as $M_\mathrm{init}$ vs. Location within HZ, for the sans basal magma ocean simulations of an M8 host star and $C_\mathrm{sat}=$ 0.1, 0.01, and 0.001. Decreasing $C_\mathrm{sat}$ leads to earlier degassing and less total final water: less water becomes sequestered in the mantle and more water is lost to space. As expected, there is slightly more total water at the end of simulations with a basal magma ocean than simulations sans basal magma ocean, but often water becomes trapped in the mantle while the surface becomes desiccated through loss to space. 

Although energy-limited loss is ongoing for the entire lifetime of the surface magma ocean for $C_\mathrm{sat} = 0.001$, the atmosphere is able to grow near the end of surface magma ocean phase in some regions of parameter space. Fig.~\ref{fig:grid_m8} shows that coupled magma ocean and deep-water cycling can maintain habitable surface conditions on a planet that would naively be expected to become desiccated based solely on the expected atmospheric loss; the planets rescued from desiccation by our model are indicated by purple boxes. This coupling may be a useful mechanism for the closely-orbiting planets of the TRAPPIST-1 system \citep{gillon17}, specifically TRAPPIST-1d (e.g., \citealt{krissansen22}).

For the highest initial water inventory simulations --- $M_\mathrm{init}=$ 50, 100, 200, and 400 Earth Oceans --- only a maximum of ${\sim}$6 Earth Oceans becomes locked in the solid mantle, which is still far below our adopted mantle saturation limit of 12 Earth Oceans. This is likely because of our chosen basal magma ocean partition coefficient of $D=0.2$; it is conceivable that, for huge $M_\mathrm{init}$ or higher $D$, the mantle could become saturated and the BMO would effectively inject water directly to the surface, potentially counteracting the loss of water to space depending on the rates of injection and loss. Although we do not see this behaviour in our model, there are regions of parameter space that allow the atmosphere to grow during deep-water cycling, in particular when the basal magma ocean lifetime is short and the injection rate is high (see, e.g., Fig.~\ref{fig:MO_cycling_6TO_RG_M8_diffBMOs}).

For simulations assuming energy-limited loss to space throughout, the loss of water to space is predictably much greater than when loss is diffusion-limited during deep-water cycling. Loss to space is especially devastating for simulations with $C_\mathrm{sat}=$ 0.001 for which an atmosphere is present and the MO is saturated throughout the entirety of the MO phase. For comparison, for both sans BMO and BMO simulations with $C_\mathrm{sat}=$ 0.01 and energy-limited loss only, 3/12 simulations around an M8 star end in a habitable surface regime compared to 9/12 with our our nominal energy-limited loss during MO, diffusion-limited loss during deep-water cycling prescription shown in the middle panel of Fig.~\ref{fig:grid_m8}; for $C_\mathrm{sat}=$ 0.001, this is further reduced from 5/12 (bottom of Fig.~\ref{fig:grid_m8}) to 2/12. These excessive amounts of energy-limited water loss are more comparable to the results of \citet{luger15} when they adopted energy-limited loss for the entirety of their simulations.

\section{Discussion}\label{sec:discussion}

The majority of our basal magma ocean simulations lead to a similar result: substantial water becomes trapped in the solid mantle --- an increasing amount with increasing basal magma ocean lifetime due to slower injection --- which can lead to dry, or even desiccated, surfaces but water-rich mantles. This means that, generally, the surface water inventory simply decreases with time as water is lost to space. 

However, there are regions of parameter space where the surface inventory is able to grow following surface magma ocean solidification. This can occur when the basal magma ocean lifetime is short enough that the water injection rate is high and the mantle is still hot and actively degassing to the surface. Regardless, this does not occur over the entirety of the parameter space, and thus this mechanism --- effectively injecting water from the basal magma ocean to the surface --- may be unlikely, especially if basal magma oceans persist for a few Gyr. Hence, the lifetime of a basal magma ocean is critical to the fate of water partitioning on an M-Earth. Based on this study, however, it appears that a basal magma ocean may not actually benefit habitability save for a few specific regions of parameter space.

A key assumption of the present study is that the orders-of-magnitude higher energy-limited escape only occurs during the earliest stage of the M-Earth, the concurrent surface magma ocean/runaway greenhouse phase, a time during which water could be mostly dissolved and protected depending on the water saturation limit of the magma. The loss of water to space is greatest at the inner edge of the habitable zone, and around late M-dwarfs.

Based on our simulation results, it seems that \citet{luger15} may have overestimated the loss of water to space when assuming energy-limited loss throughout; our energy-limited-only simulations exhibit catastrophic loss rates similar to \citet{luger15}, even with our tested $\epsilon_\mathrm{XUV}=0.1$ lower than their tested range of 0.15--0.3, while our nominal loss prescription (energy-limited during surface magma ocean and predominantly diffusion-limited otherwise) --- coupled with magma oceans and a deep-water cycle --- indicate more modest amounts of water loss, much less than the maximum expected amounts in Table \ref{tab:hzrgmaxwaterloss}. Recent studies of the TRAPPIST-1 planets find more modest loss rates of water --- roughly 1-10s of Earth Oceans, depending on the initial water content of the planet --- with less water lost from the further-out planets \citep{bolmont17, barth21}. If the surface magma ocean behaves in the way we have modelled, and if our prescription for atmospheric loss is correct, then we have done a thorough exploration of parameter space.

Another key question involves our treatment of atmospheric loss, and whether either of our loss prescriptions are realistic. For example, how well are we capturing diffusion-limited escape, which we presume predominantly occurs when the surface is solid? Perhaps a more realistic parameterization (i.e., one in which we calculate $T_\mathrm{therm}$ instead of holding it fixed, or include a background atmosphere such as the build-up of oxygen through water loss) would change the overall amount of water lost.

Moreover, the tail end of the surface magma ocean/basal magma ocean periods are the most important, especially their details; indeed, increasing $C_\mathrm{sat}$ leads to later degassing of an atmosphere from which water can be lost to space. Based on our energy-limited-only simulations, it appears that if energy-limited loss persists long after the surface magma ocean/runaway greenhouse period, a basal magma ocean could potentially be the saviour for habitability if injection continues once the host M-dwarf becomes less active. The treatment of atmospheric loss is crucial in determining when and how much water is lost from an M-Earth.

Our treatment of surface temperature $T_\mathrm{surf}$ is effectively a step function: $T_\mathrm{surf}=1800$ K during runaway greenhouse, and $T_\mathrm{surf}=293.15$ K otherwise. However, since neither the energy-limited nor diffusion-limited loss rates explicitly depend on surface temperature, the main impact will be on the thermal evolution of the mantle, which depends on the temperature contrast between the surface and the mantle, and the cycling rates, which depend on the mantle temperature. Because of this, we test two additional fixed temperate surface temperatures: $T_\mathrm{surf}=273.15$ K and $T_\mathrm{surf}=313.15$ K. Since the mantle is still an order-of-magnitude hotter than the surface, there is little effect: we find no discernible differences in the temporal plots or parameter space plots (for sans BMO simulations around an M8 host star, initiated with 6 Earth Oceans at the Inner HZ, and $C_\mathrm{sat}=0.01$). Hence, a step-wise $T_\mathrm{surf}$ seems suitable for the present study.

Early stage variables of our model are critical in determining the outcome and partitioning of water on an M-Earth, whether related to the host star or the planet itself. For example, we hold both the XUV saturation timescale and XUV absorption efficiency constant regardless of host star or orbital distance; the former may vary from 100s of Myr to multiple Gyrs and governs the energy-limited loss rate, while the latter, if decreased, would also decrease the energy-limited loss rate. Although we adopt a fixed $\epsilon_\mathrm{XUV}=0.1$, which falls roughly in the middle of the literature range, adopting $\epsilon_\mathrm{XUV}=0.01$ for a pure water atmosphere as suggested by recent studies (e.g., \citealt{ercolano10, lopez17}) would reduce the amount of water lost through energy-limited escape by an order-of-magnitude and improve water retention, potentially making the magma ocean stage --- during which water is hidden from high rates of energy-limited loss --- moot. 

As a sensitivity analysis check, we run the parameter space simulation shown in Fig.~\ref{fig:grid_m8} using a lower $\epsilon_\mathrm{XUV}=0.01$. This order-of-magnitude lower energy-limited escape results in all simulated planets maintaining water for the 5 Gyr simulations. If energy-limited escape only has an efficiency of $\epsilon_\mathrm{XUV}=0.01$ for terrestrial planets orbiting M-dwarfs, then these planets should often be habitable, in contrast with the dire predictions of \citet{luger15}.

We also assume that the surface magma ocean solidifies on the same timescale as the runaway greenhouse duration, $\tau_\mathrm{MO} = \tau_\mathrm{RG}$, since the very high runaway greenhouse temperatures should maintain a molten silicate surface and slow magma ocean cooling. This seems like a reasonable assumption, and could be further improved by linking the runaway greenhouse duration directly to the water inventory of an M-Earth, but it would not change our results: once all of the water has been lost from a planet, the rate of atmospheric loss is irrelevant.

The water partition coefficient, $D$, governs the partitioning of water between atmosphere/solid mantle following surface solidification, and the amount locked within the solid mantle is either very low (sans BMO, $D=0.001$) or moderate (with BMO, $D=0.2$); however, as previously mentioned, the latter may lead to a substantial portion of the final water trapped in the mantle by 5 Gyr. This could be made more realistic with a higher-complexity model of the surface magma ocean phase including a pressure-dependent $D$ that changes as the solidification front moves up towards the surface (see, e.g., \citealt{papale97, papale99}). 

Another important early-stage variable is the water saturation limit of the surface magma ocean, $C_\mathrm{sat}$. Based on our simulation results, we find that for higher saturation limits, water is protected against loss to space during the earliest stages of M-Earth evolution, the simultaneous surface magma ocean/runaway greenhouse stage, due to its very high solubility within the silicate melt. For lower $C_\mathrm{sat}$ --- especially when the initial dissolved water inventory is greater than the saturation limit of the magma --- an atmosphere is degassed much earlier and sometimes immediately, and thus loss persists at the energy-limited rate throughout this early stage. Indeed, the dissolution of water within a surface magma ocean protects it against the most significant energy-limited loss to space, and this protection persists longer with increasing $C_\mathrm{sat}$.

Many nuances of the habitability of terrestrial planets around M-dwarfs are not included in our box model. For example, we hold the orbital distance fixed, although potential orbital migration could move planets into, or out of, the habitable zone during their evolution. Further, although we assume a pure water vapour atmosphere throughout our simulations, the atmospheric composition will actually change over time as water molecules are photodissociated into their components; although the lighter H will predominantly be lost, the heavier O could potentially build up in the atmosphere, becoming a significant component and slowing the loss of H to space. Atomic cooling of the atmosphere may also suppress atmospheric escape rates. Further, \citet{johnstone19} note that the energy-limited escape formalism may be inappropriate for early atmospheres (primarily composed of H/He) which lack the necessary molecules to absorb XUV radiation, such as ozone within the Earth's atmosphere. Although our box model results are robust, they are merely the tip-of-the-iceberg in investigating the complex and highly-coupled nature of planetary habitability.

\section{Conclusions}\label{sec:conclusions}

The orbital distance of a terrestrial planet around its host M-dwarf will determine the duration of the runaway greenhouse phase. In principle, M-Earths could lose 3-11 Earth Oceans if energy-limited loss operates only during the runaway greenhouse; the loss could be up to 20 Earth Oceans if energy-limited loss operates for the first 5 Gyr of the M-Earth's lifetime. These amounts are modest compared to the expected water inventories of M-Earths, and through coupling interior-atmosphere water evolution, we find that desiccation of these planets may have been overstated in previous studies.

The runaway greenhouse is coeval with the surface magma ocean phase, during which most water is safely hidden below the surface; hence, the actual loss rates are even lower than the values stated above (unless energy-limited loss operates for the entire planetary lifetime, which seems unlikely). Further, if an order-of-magnitude lower XUV absorption efficiency is assumed, as suggested by recent literature, the amount of water lost is extremely small and planets remain habitable throughout the simulations. Again, the desiccation problem appears to have been overstated. 

Our results indicate that a basal magma ocean helps keep the mantle hydrated at late times, and hence could help keep planets geologically active, but it is unlikely to help replenish surface liquid water (once again, barring the strange case of energy-limited loss throughout). In general, water in the system tends to migrate into the solid mantle as it cools, which means that basal magma oceans and other deep water reservoirs will help hold onto water, but not easily bring that water up to the surface. Hence, to first-order, little to no water is lost during the magma ocean stage, and only a modest amount is lost in the remaining Gyrs, regardless of the deep-water cycle and the presence of basal magma oceans. 

If this result holds, it would bode well for the habitability of M-Earths --- most should have surface water even if they do not have operating plate tectonics or a basal magma ocean. Future studies should revisit the atmospheric loss of M-Earths using higher-complexity models (e.g., \citealt{krissansen22, lichtenberg22}), and observers should empirically establish whether any M-Earths have atmospheric water vapour. If M-Earths do not have water, it might suggest that they either form drier than expected, magma oceans have a smaller water capacity, or that atmospheric loss is more efficient than we believe. 

\section*{Acknowledgements}

KM and NBC acknowledge Colin Goldblatt for useful discussions and critical input on this study and manuscript. KM thanks Yi Huang for conversations about stratospheric moisture and convection, and Lena Noack and Tim Lichtenberg for valuable discussions about the scientific background and model results. NBC acknowledges insightful discussions with Joe O'Rourke, Leslie Rogers, and Chen Sun at the Signature of Life in the Universe Scialog workshop. KM acknowledges support from
a McGill University Dr. Richard H. Tomlinson Doctoral Fellowship,
and from the Natural Sciences and Engineering Research Council of
Canada (NSERC) Postgraduate Scholarships-Doctoral (PGS D) Fellowship.

\section*{Data Availability}

The simulation results presented within this manuscript are available in the following Zenodo repository: 10.5281/zenodo.8334934. The model is available from the corresponding author at reasonable request.


\bibliographystyle{mnras}

\bsp	
\label{lastpage}

\end{document}